\documentclass[]{spie}  %>>> use for US letter paper

\usepackage[normalem]{ulem}
\usepackage{amsmath,amsfonts,amssymb}
\usepackage{graphicx}
\usepackage[colorlinks=true, allcolors=blue]{hyperref}

\newcommand{\fourier}{\mathcal{F}}

\title{Experimental Trials With The Optical Differentiation Wavefront Sensor For  Extended Objects.}

\author[a]{Meghan Farris O'Brien}
\author[a]{Sebastiaan Y. Haffert}
\author[a]{Joseph D. Long}
\author[b]{Lauren Schatz}
\author[a]{Jared R. Males}
\author[a]{Kyle Van Gorkom}
\author[a]{Alex Rodack}

\affil[a]{University of Arizona Center for Adaptive Optics, 933 N Cherry Ave, Tucson, AZ  United States}
\affil[b]{Kirtland Air Force Base, Air Force Research Laboratory, Albuquerque, NM, USA}

\authorinfo{Further author information: (Send correspondence to S. Y. Haffert) \\M.F.B.: E-mail: mfobrien@arizona.edu \\S.Y.H.: E-mail: shaffert@arizona.eduu}

\begin{document} 
\maketitle

\begin{abstract}
Commonly used wavefront sensors - the Shack Hartmann wavefront sensor and the pyramid wavefront
sensor, for example - have large dynamic range or high sensitivity, trading one regime for the other. A
new type of wavefront sensor is being developed and is currently undergoing testing at the University of
Arizona’s Center for Astronomical Adaptive Optics. This sensor builds on linear optical differentiation
theory by using linear, spatially varying half-wave plates in an intermediate focal plane. These
filters, along with polarizing beam splitters, divide the beam into four pupil images, similar to those
produced by the pyramid wavefront sensor. The wavefront is then reconstructed from the local
wavefront slope information contained in these images. The ODWFS is ideally suited for wavefront
sensing on extended objects because of its large dynamic range and because it operates in a pupil plane
which allows for on chip resampling even for arbitrarily shaped sources. We have assembled the ODWFS
on a testbed using a 32 x 32 square 1000 actuator deformable mirror to introduce aberration into a
simulated telescope beam. We are currently testing the system’s spatial frequency response and are
comparing the resulting data to numerical simulations. This paper presents the results of these initial
experiments.
\end{abstract}


\section{INTRODUCTION}
\label{sec:intro}  % \label{} allows reference to this section

Atmospheric turbulence degrades the image quality of ground-based astronomical observations and limits spatial resolution to a few arcseconds. Most modern ground-based telescopes now employ adaptive optics to correct for atmospheric turbulence. Adaptive optics systems use wavefront sensors to measure aberrations across the aperture and reconstruct the wavefront. The conjugate of the reconstructed wavefront is then applied to a deformable mirror (DM) in the path of the science beam, flattening the wavefront; the effectiveness of the wavefront sensor determines the degree to which the phase conjugate cancels the aberration. The goal of an Adaptive Optics system is to correct the telescope's beam to as close to diffraction limited as possible\cite{tyson2000introduction}.

The most commonly used wavefront sensors for astronomical adaptive optics are the Shack-Hartmann wave- front sensor (SHWFS) and the Pyramid Wavefront Sensor (PWFS). While the SHWFS has been the wavefront sensor of choice throughout most of the history of astronomical Adaptive Optics, it is currently being succeeded by the PWFS in most new telescopes.
Both the SHWFS and PWFS have an inverse relationship between dynamic range and sensitivity. The SHWFS has considerable sensitivity, but saturates under large aberrations and thus has large  dynamic range. In closed loop situations, the SHWFS noise contributions are inversely proportional to the total amount of photons per sub-aperature. The PWFS maintains good signal to noise ratio for a large range of aberrations. Modulating the Pyramid increases dynamic range but decreases sensitivity. In closed loop, the PWFS noise contribution is proportional to the total amount of photons on the aperture. The PWFS offers better photon noise propagation than the SHWFS along with a higher limiting magnitude. One of the main limitations of the PWFS is that it is most effective for point sources and not as effective for extended spots like those from Laser Guide Star beacons. Although the SHWFS is somewhat effective at extended sources, it is generally not amenable to easy changes in wavefront sampling, and suffers from relatively strong aliasing errors\cite{knight2021analysis}. Other Wavefront Sensors, such as the Ingot Wavefront Sensor, have success with extended sources, but suffer from their own limitations.

The Optical Differentiation Wavefront Sensor (ODWFS) was first conceptualized in 1972 as a way to optimize the Schlieren technique of phase contrast imaging to obtain the transmittance gradient of an extended source.\cite{sprague1972quantitative} Twelve years later Bortz \& Thompson proposed several methods of phase reconstruction by varying the amplitude transmittance of a spatial filter placed in the telescope's focal plane.\cite{bortz1983phase} These methods utilized the first derivative of the pupil irradiance to reconstruct the phase. The three phase irradiance method with opposite transmittance gradient was chosen to be investigated further to combine some of the benefits of the SHWFS and inadvertently the PWFS by creating a filter that varies the amplitude by the square root of position.\cite{horwitz1994new} In the early twenty-first century the amplitude filter crucial for the ODWFS was augmented to create a SNR and dynamic range comparable to that of the SHWFS.\cite{feeney2001theory} This was a major advance, because it provided significant gain over the SHWFS in extended source WFS\cite{oti2003analysis}. Although the ODWFS at this point provided better dynamic range and SNR in extended source experiments, it only allowed 50 percent of the light to be utilized for phase reconstruction.\cite{oti2005improvements} A decade later researchers from Leiden University designed an improved linear amplitude focal plane mask using liquid crystal polymers in order to make the ODWFS nearly 100 percent photon efficient.\cite{haffert2016generalised}

Recent work\cite{haffert2016generalised, haffert2018sky} discusses wavefront sensing by spatial filtering using polarization focal plane filters, and it is this methodology that we use here. We are specifically interested in testing the suitability of the ODWFS for extended source wavefront sensing. Previous work investigated the performance for 3D objects like Laser Guide Stars (LGS)\cite{haffert2020using}. However, many extended objects can easily be approximated as two-dimensional because their surface height is negligible compared to their extent. This is not true for LGS which has a height extent that is larger than its width. In the next section of this paper, we will discuss our Experimental Setup. In the section Calibrating the ODWFS, we will explain our method of calibrating the wavefront sensor. In Experimental Design we will detail our goals and our data. In Analysis, we will present and interpret our results. In Conclusions, we will discuss the broader implications of our findings and outline our next experiments.

\section{The ODWFS}
The ODWFS consists of two lenses with a focal plane mask in the intermediate focus. The output electric field of such a setup is the Fourier transform of the focal plane filter multiplied by the Fourier transform of the input electric field. 
\[E_1 = \fourier[t\fourier[E_0]\].
Here, \(t\) represents the effective amplitude filter of which there are two created by each of the four focal plane masks. Two within the x direction and two in the y direction. In the positive and negative x-direction we have, 
\[t_+ = \sin[\frac{\pi}{4}(1+\frac{k}{k_m})]\]
\[t_- = \cos[\frac{\pi}{4}(1+\frac{k}{k_m})]\]
Where \(k\) represents the focal plane coordinate and \(k_m\) represents the normalization factor and extent of the filter in spatial frequencies. The output electric field is the convolution of the input electric field and the Fourier Transform of the filter.
\[E_\pm = E_0 * \fourier[t_\pm]\]
We are not so much interested in the electric field as in the intensity patterns in the pupils, from which we reconstruct the wavefront. The pupil intensity, 
\[I_\pm = 8\pi[I_0(x+x_0)+I_0(x-x_0)\pm 2\Re(iE_0(x+x_0)E_0(x-x_0)^t\]
Because \(E_0 = Ae^{i\phi(x,y)} \) is the electric field \(I_\pm\) becomes, \[I_\pm = 8\pi(I_0^+ \pm I_0^- \pm 2A^+A^-\sin(\phi^+-\phi^-))\]
Which leads us the basis of reconstruction, 
\[\sin^{-1}[\frac{I_+-I_-}{I_++I_-}] \approx \frac{\pi}{2}\frac{1}{\theta_0}\frac{\partial W}{\partial x}\]
Where \(\theta_0\) is the on-sky size of the focal plane mask. To extend this to extended sources we have to integrate the response over the full source.Because \(E_0 = Ae^{i\phi(x,y)} \) is the electric field \(I_\pm\) becomes, \[I_\pm = 8\pi(I_0^+ \pm I_0^- \pm 2A^+A^-\sin(\phi^+-\phi^-))\]

\section{Experimental Setup}
\begin{figure}
 	\centering
 	\includegraphics[width=1\textwidth]{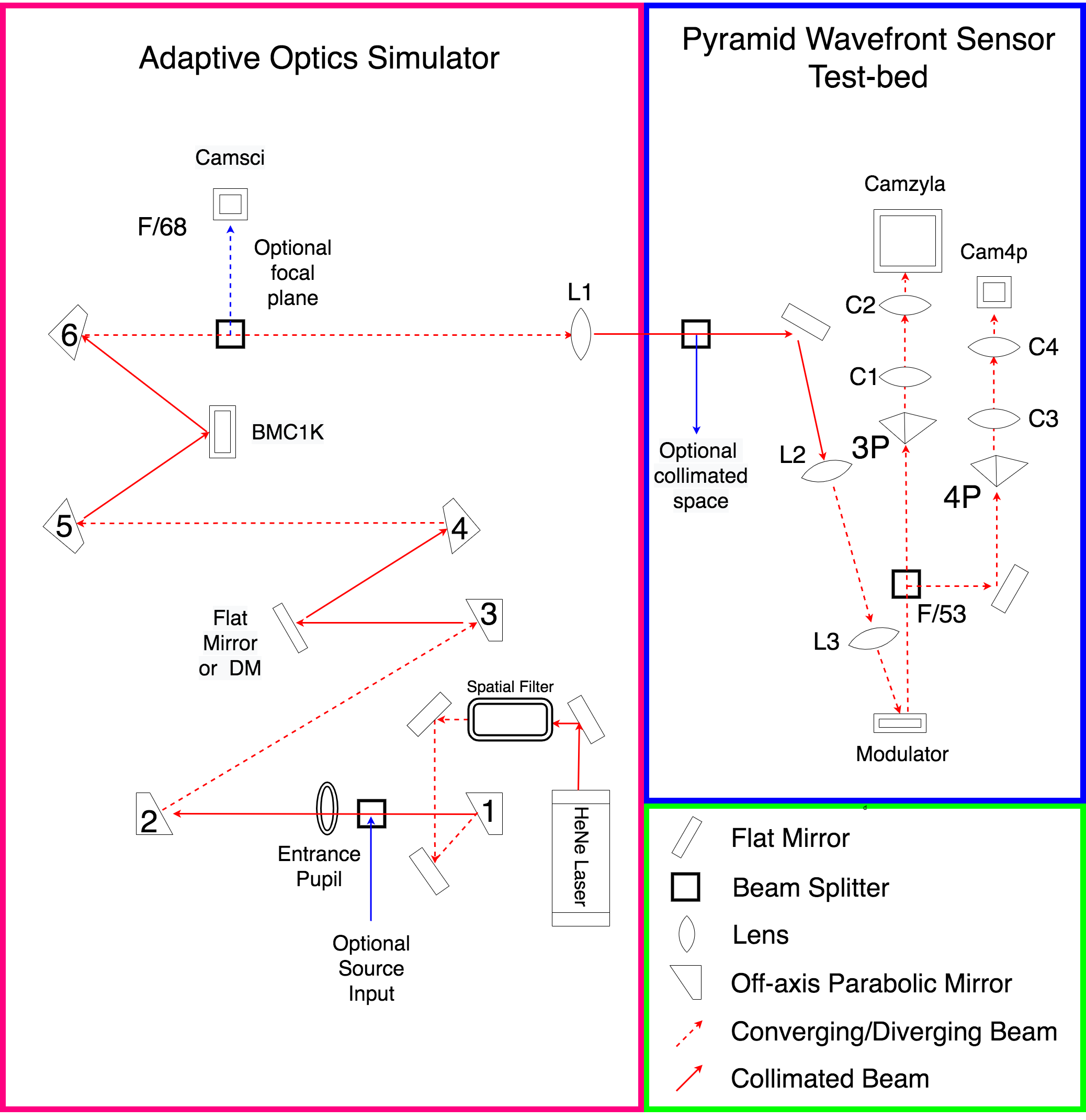}
 	\caption{ Comprehensive Adaptive Optics and Coronagraph Test Instrument. The AO simulator and PWFS. The light is passed to the PWFS testbed which includes a modulation mirror, a 4PWFS and a 3PWFS. }
 	\label{fig:Comprehensive Adaptive Optics and Coronagraph Test Instrument}
 \end{figure}

The ODWFS has been implemented within a larger experimental AO testbed, the Comprehensive Adaptive Optics and Coronagraph Test Instrument (CACTI). CACTI system is composed of two major systems: an adaptive optics simulator, and a PWFS testbed
composed of both a three and a four-sided pyramid wavefront sensor\cite{schatz2020first,schatz2021three}. A schematic of this system is shown in Figure \ref{fig:Comprehensive Adaptive Optics and Coronagraph Test Instrument}.

\begin{figure}
 	\centering
 	\includegraphics[width=1\textwidth]{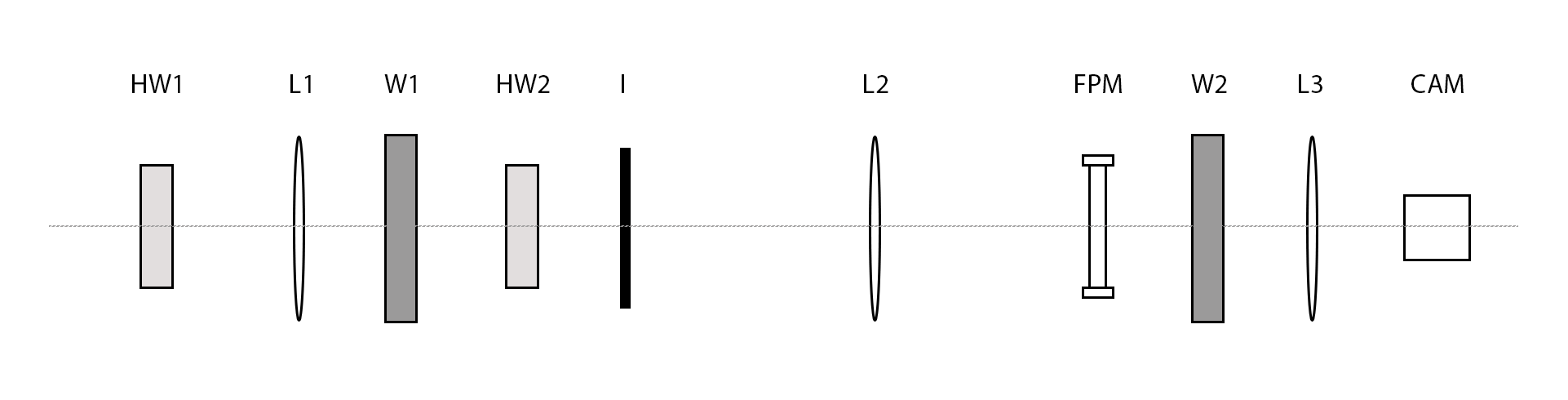}
 	\caption{ A schematic of the ODWFS layout. The components of the ODWFS consist of a half-wave plate (HW1) followed by a lens that collimates the beam (L1). The beam is then split into two by a Wollaston prism (W1) and the outgoing polarization is rotated by a second half-wave plate (HW2). The second lens (L2) is used to focus the beam onto the focal plane mask (FPM). After the FPM, a second Wollaston prism (W2) splits the beams again into two resulting in four output beams. These four beams are collimated by a third lens (L3). }
 	\label{fig:odwfs_schematic}
 \end{figure}

\begin{figure}
 	\centering
 	\includegraphics[width=1\textwidth]{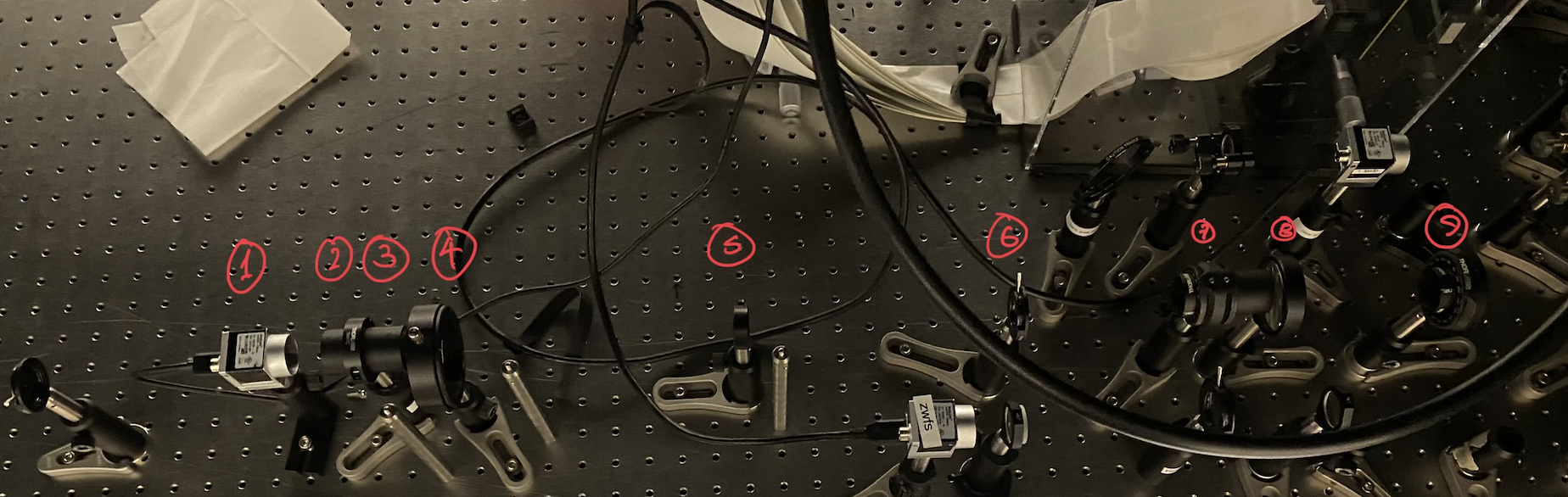}
 	\caption{ Photograph of the ODWFS layout. Labeled as follows; 1.) Science Camera 2.) 180mm Lens 3.) Second Wollaston beam splitter 4.) Linear Amplitude Transmittance filter (FPM) 5.) 75mm lens 6.) Iris 7.) Second Half Wave Plate and First Wollaston Prism  8.) 200 mm lens 9.) First Half-wave plate. }
 	\label{fig:odwfs_labphoto}
 \end{figure}

The beam source for CACTI is a 633 nm Helium-Neon laser which is spatially filtered through a 10 micron pinhole. Test aberrations are introduced into the beam, and corrected, by CACTI’s 1024 actuator Boston Micromachine 1K (BMC1K) deformable mirror. The deformable mirror was flattened using an interferometer. The beam is split off after Off-axis parabola (OAP) 6, where we added an additional 50/50 beamsplitter. The transmitted part is sent to an imaging camera that monitors the PSF and the reflected part is sent to the ODWFS setup. A schematic of the ODWFS setup is shown in Figure \ref{fig:odwfs_schematic}. Figure \ref{fig:odwfs_labphoto} shows a picture of the assembled ODWFS in CACTI. The beam arrives at the ODWFS and first strikes a halfwave plate (HW1 in figure 2), which serves to rotate the vertical polarization of the HeNe beam by 45 degrees to achieve the desired entry polarization for the Wollaston prism. This would not be necessary for unpolarized starlight. The laser source we use is linearly polarized, that creates an uneven power split after the first Wollaston prism. The half-wave plate is used to create an even power split between the two beams.

%We have added a The optical layout we have utilized to test the response of the ODWFS is shown in Figure \ref{fig:odwfs_schematic}. with element. references is shown in figure 2, and a summary of pertinent design specifications is presented in table 1.  

The beam then passes through a 200  mm lens (L1), which collimates the beam. The Wollaston prism at W1 splits the beam at an angle of 1.33$^\circ$ degrees. A second half-wave plate rotates the polarization by 45 degrees to create the correct angle of polarization for the focal plane mask. The polarization states of these beamlets are now at 45 and -45 degrees (diagonal and anti-diagonal states). Both of these now pass through the iris at position I, which was used to align the lenses before the polarization optics were placed in the beam. A 75 mm lens (L2) brings the beam to a focus, such that its focal point falls on the focal plane mask (FPM). The beam then diverges and passes through a second Wollaston prism (W2), which splits both beamlets into two, resulting in four outgoing beams. Each pair of beams consist of perpendicular polarization states. A 180 mm lens at L3 recollimates the divergent beams, which now form four pupil images on the Basler camera located at position CAM.
The focal plane mask contains a linear amplitude filter in a focal plane filtering setup. This can be thought of as a
continuous Foucault knife edge test instead of the normal discrete knife edge test. 

\section{Lab calibration}
Calibration of the ODWFS occurs in two steps: one, optimization of the Strehl ratio of the PSF, and two, calibration of the deformable mirror and the wavefront sensor with respect to each other.

Both steps are done using code written in Python and executed through Jupyter notebooks on the control computer. Optimization of the Strehl ratio involves scanning through a range of amplitudes applied to each of first 30 Zernike modes. We start with the first mode, and cycle through a series of amplitudes; for each applied amplitude, the peak brightness of the PSF is recorded. A quadratic fit is applied to the data to determine the amplitude that maximizes the PSF. From this, a best correction is determined, and then this is applied before the measurements of the next Zernike mode in the sequence. As this is inherently a non-linear optimization process, we iterate several times over the same set of Zernike modes. Optimization is considered complete after we have applied this process to the first 30 Zernike modes.

The presence of non-common path aberrations (NCPAs) inherent in our system design present a challenge. The NCPA causes the response of the ODWFS to be non-zero around the PSF optimized point. We address this by averaging 100 measurements at the optimized DM shape to estimate the zero point. The reconstructed wavefront coefficients are then subtracted to correct for the offset. The DM and ODWFS are calibrated with respect to each other by measuring the response matrix. The response matrix was measured by the push-pull method. In a push-pull calibration, we add a Zernike mode with a positive amplitude to the DM ('push'), and then we take another measurement but with a negative amplitude ('pull'). The difference between the push and pull measurements gives us a single column of the response matrix. This is then repeated for each mode that we want to measure, which in the end creates the full response matrix. The response matrix transforms modal coefficients into a measurement. We usually have a measurement for which we want to know the incoming wavefront. This can be done with the pseudo-inverse of the response matrix, which we will call the reconstruction matrix. The measurement process has three steps, create the normalized differences and extract the slope signals. Use the reconstruction matrix and the slope signals to calculate the modal coefficients. And as a last step, we subtract the zero point offset. All calibrations are done with a point source. This is done because we are investigating how the performance degrades when there is a difference between the calibration source size and the actual source size. Most lab-based calibration are done with point sources, which is why we have chosen the point source as the calibration source.

We did not have access to an extended source that is large enough for our experiments at the current stage of the project. That is why an extended source was simulated with the DM. The DM applied a sequence of tip/tilt that traced the source shape. We took an image after each new tip/tilt position and averaged all images to create the incoherent extended object. The sequential image approach avoids the need to synchronize exposure times and the start of the DM sequence.

\section{Data and Analysis}
The performance of the ODWFS for extended objects was tested by two different experiments. Both experiements tests the response of the ODWFS as function of different input shapes. We used the DM to create circular extended objects with different diameters. The diameters that we used were 0, 1, 2, 3 and 4 $\lambda/D$. It was not possible to make larger diameters due to the limited stroke on the DM.

 \begin{figure}
 	\centering
 	\includegraphics[width=1\textwidth]{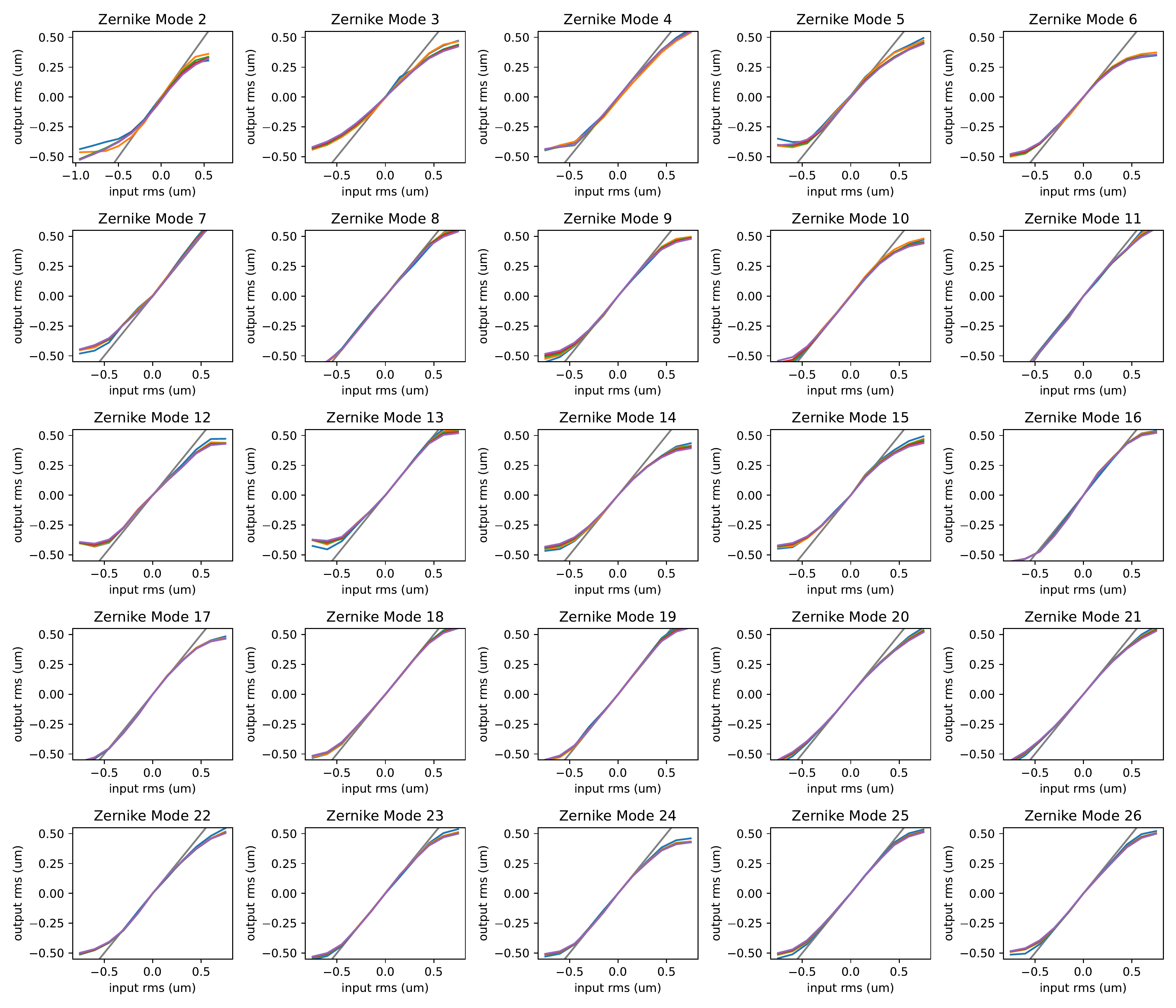}
 	\caption{ The input amplitude of each mode verus the reconstructed amplitude. The different colors in each figure represents a different source diameter. The curves seem to be identical and independent of the tested source diameters. Only the tip/tilt curves show variation, which could be due to slow drifts of the PSF.}
 	\label{fig:dr_odwfs}
 \end{figure}

The first experiment was done to explore how the linear range of the ODWFS changes when the diameter is increased. For each diameter we went applied a range of different amplitudes for a particular Zernike mode. The results for the first 25 Zernike modes (excluding piston) are shown in Figure \ref{fig:dr_odwfs}. The results show that there is nearly no dependency on the size of the source. Only the tip and tilt modes show some variation, which could be due to slow drifts of the PSF. The higher order modes show identical response. These experiments demonstrate that the response is independent of source sizes up to $4 \lambda/D$.

\begin{figure}
 	\centering
 	\includegraphics{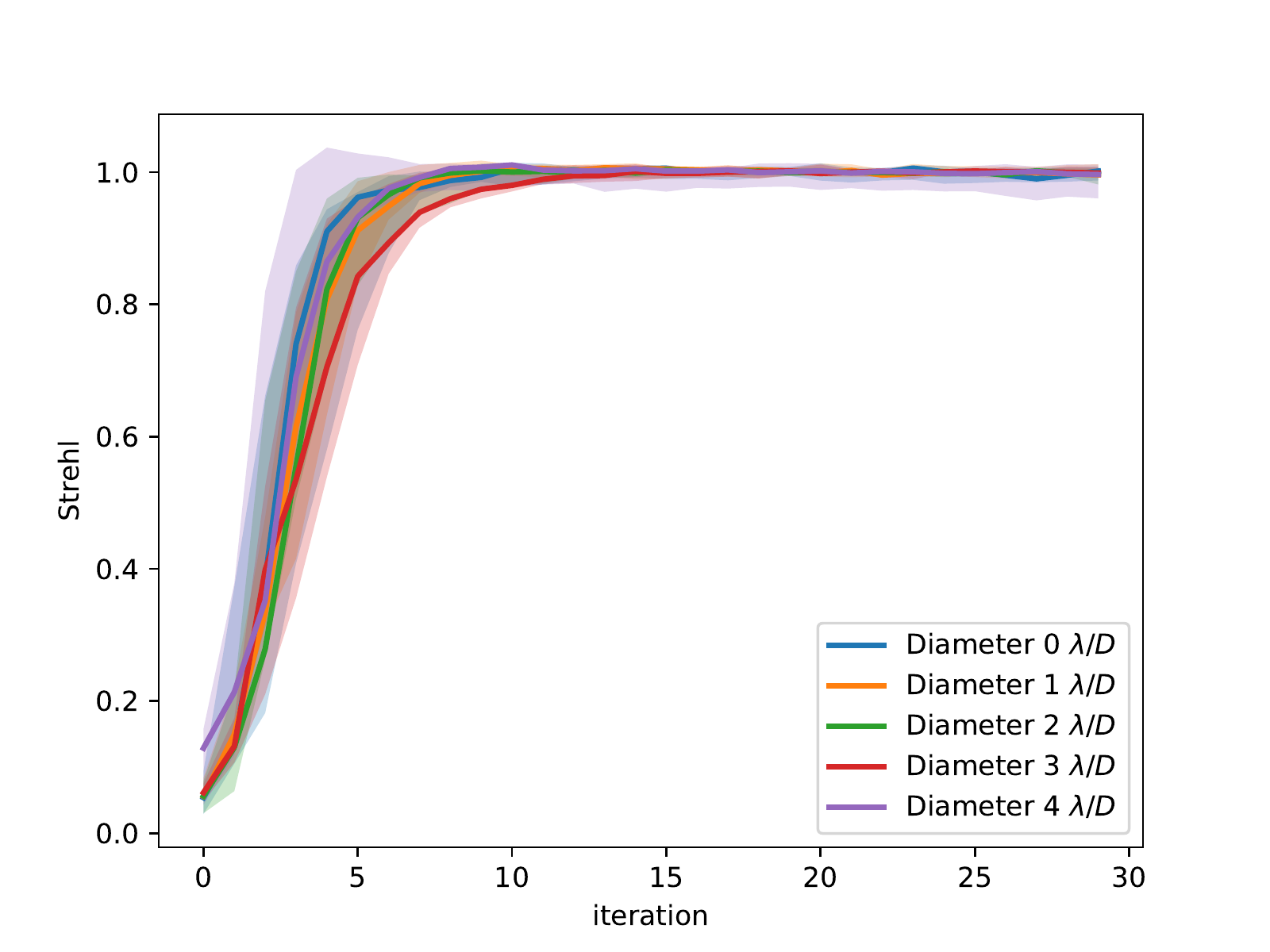}
 	\caption{The closed-loop Strehl as function of the number of iterations. Each color represents a different source diameter. The solid lines show the median performance over 5 trials and the shaded area represents the minimum and maximum Strehl in the 5 trials. All curves except for the 3$\lambda/D$ diameter  overlap.}
 	\label{fig:Strehl Ratio Response}
 \end{figure}

In the second experiment we applied a random deformation on the DM and tried to correct this in closed-loop. This experiment was repeated 5 times for each diameter. The Strehl versus iteration is shown in Figure \ref{fig:Strehl Ratio Response}. The closed-loop results show that almost all diameters have equal performance and, all of them fall within the envelop of the point source ($0 \lambda/D$) curve. The $3 \lambda/D$ diameter curve has a slower convergence rate than the other diameters. It is not clear why this curve is different because the 4 $\lambda/D$ diameter curve follows the point source curve again. This could be a statistical effect due to the low number of trials per curve.

\section{Conclusion}
We have reported the implementation of the ODWFS within the CACTI testbed at the University of Arizona. This experiment is to test the performance of wavefront sensing and control on extended objects. The current ODWFS is aligned and performing as expected. The DM in the setup is used to create extended objects with diameters up to 4 $\lambda/D$. Our experiments demonstrate that there is no degradation in performance for extended objects (up to 4 $\lambda/D$) in our setup. This has been tested by measuring the linear range for the first 25 Zernike Modes for each source diameter. We also ran several trials for each source diameter where we corrected a random deformation on the DM in closed-loop. The results show no difference in performance for the tested range of source diameters.

In future work, we will explore larger source diameters because the linear approximation will start to fail when the source size approaches the size of the focal plane mask. The current experiments show that the ODWFS has an excellent linear range and can handle resolved objects without any problems. This makes the ODWFS a competitive wavefront sensor to the Shack-Hartmann wavefront sensor for extended objects.

%The MagAO-X lab has been experimenting with ODWFS. It is the wavefront sensor component of the CACTI experiment we have shown how the ODWFS has been calibrated and briefly discussed the Fourier trials and our experiments with extended light sources. The results of these initial experiments show that the ODWFS is an ideal candidate for continued testing for extended source wavefront reconstruction. 

\acknowledgments % equivalent to \section*{ACKNOWLEDGMENTS}
Support for this work was provided by NASA through the NASA Hubble Fellowship grant \#HST-HF2-51436.001-A awarded by the Space Telescope Science Institute, which is operated by the Association of Universities for Research in Astronomy, Incorporated, under NASA contract NAS5-26555. The authors would also like to thank the University of Arizona's Postdoctoral Research Grants for providing funding for this project. 

% References
\bibliography{report} % bibliography data in report.bib
\bibliographystyle{spiebib} % makes bibtex use spiebib.bst

\end{document}